**Title:** Import and export of horticultural products in Portugal

**Author:** Vítor João Pereira Domingues Martinho

Unidade de I&D do Instituto Politécnico de Viseu

Av. Cor. José Maria Vale de Andrade

Campus Politécnico

3504 - 510 Viseu

PORTUGAL

e-mail: vdmartinho@esav.ipv.pt

# Import and export of horticultural products in Portugal

**Abstract**

With this work it is analyzed the import and export of horticultural products between Portugal and the other world countries. It is used data about Portuguese international trade of vegetables from 2006 to 2010. The data were obtained from the INE (Statistics Portugal), gently given by the AICEP (Trade & Investment Agency). It is did some estimations taking into account the models from the convergence theory, with panel data and using methods by fixed effects, random effects and dynamic effects, for the Portuguese import and export of vegetables, separately. It is found convergence in all estimations. The volatility was also tested. All the tests show no stationary of the data. So, in statically means the data show weak regularity. In this way all the conclusion, must be did very carefully. This lack of regularity is a result of lack of a national coherent policy for the sector.

**Keyword:** Horticultural products, international trade, data analysis.

## 1. Introduction

The horticultural sector is an important economic activity for Portugal, because the natural conditions to this production. Sometimes the framers in their decisions do not opt for these productions, because are influenced by the European and national agricultural policies.

Having regard to the Horticulture Survey 2000 (IH2000) there are 21725 farms in Portugal Continental with horticultural products, which in terms of area corresponds to 31070 hectares. These values should be around about ten percent of the total number of farms in Portugal and of the utilized agricultural area. The production area of vegetables in Portugal is relatively fragmented, although the area par excellence is in the center and south. This shows the relative importance of horticultural products, but also shows the huge potential that still exists in these productions. The extensive horticulture assumes some importance in Portugal and includes most of the species used for industrial processing, especially tomatoes, but also peppers, cabbage, broccoli, among others. This extensive horticulture occupies about two thirds of the area for the crops, being held on 15316 farms. The production of vegetables in an intensive, occupying the second position in terms of area and are held in 11990 farms, representing about 30% of the area used in horticulture. The intensive horticultural in glasshouse are carried out in 3175 farms. These productions are therefore of minor importance in terms of area, since it represents only 4% of the horticultural (PPG 2007).

In regional terms, the extensive horticultural crops appear namely in the Ribatejo and Oeste, which owns 66% of the area and 32% of the farms of Portugal, followed by Alentejo with 21% of the area and 11% of farms. These zones have the best conditions in Portugal Continental for these productions, because have the most important requirements, namely sun and water The intensive horticultural are as well mostly made in Ribatejo and Oeste (72% of the area and 38% of the farms of the Continent), although some take importance in Entre Douro e Minho and Beira Litoral which together represent 18% of area and 36% of farms. Despite being in Entre Douro e Minho which has the highest number of farms with intensive horticultural in greenhouse, about 35% of the continent, are the regions of Ribatejo and Algarve that have the largest area with this utilization, assuming that latter region the first position with 42% of the total area of greenhouses. Again, the factors temperature and the water having, here, an important role to play in the horticultural productions, considering the importance of these two factors in the Algarve region (GPP, 2007).

In this context seems important to analyze the import and export of horticultural products in Portugal. This is because it is believed that the international trade is an important indicator about the dynamic of the sectors.

In this work the most important is analyze the Portuguese international policy for this sector. Try understanding if there is an objective policy to the international trade of vegetables and from these conclusions try understanding the adequacy of the global policies for the horticulture in Portugal.

For this it is analyzed the data about the Portuguese international trade for vegetables and made some econometric analysis.

## 2. Review of the literature

International trade of horticultural products is progressively more important in many regions of the world, but relatively little studies have considered the import evolution of the horticultural crops (Rickard et al., 2009). In fact, there are few works about the international trade of products from the horticultural sector. Despite the little number of papers existents concerning the import and export of vegetables, some are about the influence of several factors in the international trade, others about the importance of the sector in developing countries, others about the liberalization consequences and others, yet, about the quality standards.

With data among 1991 and 2005, Rickard et al. (2009), discover which import demand for horticultural products in developed countries has been determined firstly by prices and the level of trade liberalization. Generally the price import is related with the liberalization, or not, in the international trade of each country. On other side income and diet attributes were more significant in emerging countries. These means which when the level of live improve the preoccupations are different, as we know from the economic theory. In addition, their results demonstrate that the causes of import demand differ across the chosen crops, and consequently conclusions can be lost if data for horticultural products are aggregated. These are interesting conclusions for international operator in these productions and contributions for the international trade policies makers, because from here it was identified some factors which influence the horticultural trade in different countries.

On other hand, it must be taken into account the specificity of each country at the level of the productive structure. Because some countries are capital intensive and others are labor intensive. For example, it is, already, recognized that the growth and development social and economic of developing countries will be more encouraged with the export of vegetables, considering which these countries have more labor than capital or land. So, these countries have a comparative advantage in intensive labor productions, as the horticultural sector as compared to other production. The past tendency has also revealed an enlarge in market share of developing countries not only in the international trade of vegetables but also in terms of share of vegetables in their total agricultural exports (Shah, 2007a).

In recent years, considering the quickly changing in the international market circumstances, horticultural sector between diverse agricultural sectors have obtained much interest in international trade. The situation for the openness of trade in the horticultural productions may be related with the triumph of the international discussions about the trade questions, namely in terms of Uruguay Round conversations and in terms of the World Trade Organization. The consequent importance of the horticulture in the world trade of agricultural productions and namely in the developing countries was referred by several authors (Shah, 2007b).

Understand the social context of each country and the history of each society is important to explain the economic evolution of the several sectors, namely the agricultural sectors and the horticultural productions. This is what Freidberg (2003) did for understand the transformations in the actual fresh vegetables trades pattern among Africa and the Europe. In the 1990s first British and later French supermarkets created codes defining conditions for the food safety, namely agricultural best practice and, in the UK, ethical trade. Consequently, these conditions represent commercial contributions not only to guarantee but also return from an increasingly preoccupation with the food safety. These conditions also assure to induce changes in the international retailers of fresh produce supply chains, but, as expected, not homogeneously (Freidberg, 2003).

The analysis of the environment consequence is an important issue which must be considered, namely because sustainability questions. The studies show which openness in the international trade drives to an augment in total greenhouse gas emissions in about 6% considering the reference scenario value in 2015. The augment in $CO_2$ emissions are determined by a rapid increase of agricultural area, principally in South America and Southeast Asia. The results are less obvious in situations where trade openness is only implemented partially. This shows that different patterns of liberalization must be analyzed for each region and per product before try generalize conclusions on the impact of trade openness (Verburg et al., 2009).

**3. Data analysis**

The countries presented in the four following tables represent about ninety per cent of the vegetables international trade between Portugal and world.

Analyzing table 1 we find that Portugal did not import, in the period considered, vegetables from Angola and Cape Verde.

Import from Guinea-Bissau only other vegetables and from Mozambique only leguminous. From Morocco import several horticultural products, but mainly leguminous. More or less the same happen with Brazil, Canada, United States of America and China.

From Switzerland and Sao Tome and Principe Portugal import few vegetables. From Germany, Denmark, France, Holland, Ireland and Luxembourg the Portuguese import mainly potatoes.

Import from Austria, Belgium and United Kingdom mainly vegetables cooked or uncooked. From Spain, Estonia and Italy, Portugal import several horticultural products.

This context show us that the Portuguese import principally potatoes, other vegetables, leguminous and cooked or uncooked vegetables. As consequence will be important to analyze in future works what happen with the markets of these vegetables, namely about the producers and the policies in these sectors. Because, Portugal has optimal natural conditions to produce all these vegetables. Objective policies are needed to change this situation.

Analyzing, in general, the data in the table 1, it is found that there is not a regularity in the several data, what may represent a deregulation of the sector and incoherent policy to the sector and for it international trade.

**Table 1. Horticultural products, in different forms, import percentage relatively to the total of each country**

| | year | Angola | Cape Verde | Guinea-Bissau | Mozambique | Sao Tome and Principe | Morocco | Brazil | Canada | United States of America | Switzerland | China | Germany | Austria | Belgium | Denmark | Spain | Estonia | France | Holland |
|---|---|---|---|---|---|---|---|---|---|---|---|---|---|---|---|---|---|---|---|
| Potatoes, fresh or chilled | 2006 | | | | | | 19 | | 0 | | | | 13 | | 10 | 100 | 20 | 30 | 66 | 72 |
| | 2007 | | | | | | 59 | | 1 | | | | 18 | | 11 | 100 | 19 | | 70 | 74 |
| | 2008 | | | | | | | | 0 | | | | 45 | | 13 | 98 | 20 | | 66 | 69 |
| | 2009 | | | | | | | | | | | | 49 | | 12 | 100 | 15 | 25 | 73 | 71 |
| | 2010 | | | | | | 3 | | 0 | | | | 42 | | 6 | 96 | 17 | 18 | 68 | 68 |
| Tomatoes, fresh or chilled | 2006 | | | | | | 23 | 4 | | | | | 21 | | 1 | | 14 | | 1 | 1 |
| | 2007 | | | | | | | | | | | | 16 | | 0 | | 14 | | 1 | 1 |
| | 2008 | | | | | | | | | | | | 0 | | 1 | | 10 | 22 | 3 | 1 |
| | 2009 | | | | | | | | | | | | 1 | | 1 | | 15 | 28 | 2 | 1 |
| | 2010 | | | | | | 1 | | | | | | 1 | | 0 | | 14 | 33 | 1 | 0 |
| Onions, shallots and other prod vegetables, fresh or chilled | 2006 | | | | | | | | | | | 1 | 26 | | 2 | 0 | 13 | 65 | 6 | 3 |
| | 2007 | | | | 2 | 2 | | | | | | 1 | 25 | | 1 | 0 | 13 | | 9 | 5 |
| | 2008 | | | | | | | | | | | 1 | 14 | | 0 | | 12 | 21 | 8 | 1 |
| | 2009 | | | | | | | | | | | 0 | 3 | | 1 | | 15 | 24 | 3 | 1 |
| | 2010 | | | | | | | | | 2 | | 0 | 0 | | 2 | | 18 | 23 | 7 | 6 |
| Cabbages, cauliflower, cabbage or kale kohlrabi, etc.., Fresh or chilled | 2006 | | | | | | | | | | | 100 | 6 | | 1 | | 4 | | 1 | 1 |
| | 2007 | | | | | | | | | | | | 6 | | 1 | | 3 | | 2 | 1 |
| | 2008 | | | | | | | | | | | 100 | 1 | | 0 | | 2 | 1 | 1 | 2 |
| | 2009 | | | | | | | | | | | | 1 | | 1 | | 3 | 2 | 1 | 2 |
| | 2010 | | | | | | | | | | | | 0 | | 1 | | 4 | 1 | 1 | 2 |
| Lettuce and chicory, fresh or chilled | 2006 | | | | | | | | | | | | 0 | | 1 | | 1 | | 1 | 0 |
| | 2007 | | | | | | | | | | | | 0 | | 3 | | 2 | | 0 | 0 |
| | 2008 | | | | | | | | | | | | 0 | | 3 | | 2 | 0 | 0 | 0 |
| | 2009 | | | | | | | | | | | | 0 | | 2 | | 2 | 0 | 1 | 0 |
| | 2010 | | | | | | | | | | | | 0 | | 1 | | 2 | 0 | 1 | 1 |
| Carrots, turnips, salad beetroot, radishes, etc., fresh or chilled | 2006 | | | | | | | | | | | | 13 | | 6 | | 6 | | 10 | 1 |
| | 2007 | | | | | | | | | | | 0 | 11 | | 6 | | 5 | | 4 | 1 |
| | 2008 | | | | | | | | | | | 0 | 2 | | 1 | | 6 | 3 | 3 | 1 |
| | 2009 | | | | | | | | | | | | 0 | | 3 | | 5 | 0 | 3 | 0 |
| | 2010 | | | | | | | | | | | | 0 | | 7 | | 6 | | 6 | 1 |
| Cucumbers and gherkins (gherkins), fresh or chilled | 2006 | | | | | | 1 | | | | | | 0 | | 0 | | 1 | | 0 | 0 |
| | 2007 | | | | | | 0 | | | | | | 0 | | 0 | | 1 | | 0 | 0 |
| | 2008 | | | | | | 2 | | | | | | 0 | | 0 | | 1 | 2 | 0 | 0 |
| | 2009 | | | | | | 7 | | | | | | 0 | | 0 | | 1 | 1 | 0 | 0 |
| | 2010 | | | | | | 0 | | | | | | 0 | | 0 | | 1 | 1 | 0 | 0 |
| Leguminous vegetables, shelled or unshelled, fresh or chilled | 2006 | | | | | | 46 | | | | | | 3 | | 1 | | 10 | | 1 | 0 |
| | 2007 | | | | | | 38 | 0 | | | | | 2 | | 0 | | 9 | | 1 | 0 |
| | 2008 | | | | | | 71 | 0 | | | | | 0 | | 0 | | 7 | 16 | 0 | 0 |
| | 2009 | | | | | | 51 | | | | | | 0 | | 0 | | 7 | 5 | 0 | 0 |
| | 2010 | | | | | | 33 | | | | | 0 | 0 | | 0 | | 5 | 4 | 0 | 0 |
| Other vegetables, fresh or chilled | 2006 | | 99 | | 6 | 49 | | | | | | 0 | 6 | 66 | 1 | | 13 | 1 | 1 | 6 |
| | 2007 | | 100 | | 1 | 11 | | | | | | | 7 | 1 | 1 | | 16 | 100 | 1 | 6 |
| | 2008 | | 100 | | 26 | 21 | | | | | | 0 | 5 | 8 | 3 | | 18 | 16 | 3 | 7 |
| | 2009 | | 100 | | 34 | 0 | | | | | | 0 | 7 | 7 | 5 | | 14 | 13 | 2 | 8 |
| | 2010 | | 100 | | 2 | 2 | | | | 0 | | 0 | 4 | 3 | 3 | 1 | 15 | 16 | 1 | 9 |
| Vegetables, uncooked or cooked by steaming, frozen | 2006 | | | | | | | | | | | | 3 | 3 | 34 | 73 | 9 | 4 | 9 | 7 |
| | 2007 | | | | | | | | | | | | 3 | 3 | 99 | 72 | 10 | | 9 | 7 |
| | 2008 | | | | | | | | | | | | 1 | 8 | 92 | 77 | 11 | 0 | 12 | 11 |
| | 2009 | | | | 1 | | | | | | | | 3 | 18 | 93 | 71 | 11 | 0 | 10 | 12 |
| | 2010 | | | | | | | | | | | 100 | 1 | 15 | 78 | 72 | 10 | 0 | 11 | 9 |
| Production vegetables preserved, but not for food in this state | 2006 | | | | | | | | | | | | 1 | 1 | | 0 | 0 | 5 | | 0 | 0 |
| | 2007 | | | | | | | | | 0 | | | 2 | 1 | | | | 4 | | 0 | 0 |
| | 2008 | | | | | | | | | 0 | | | 2 | 3 | | 0 | | 6 | | 0 | 0 |
| | 2009 | | | | | | | | | 0 | | | 3 | 2 | | | 0 | 4 | 0 | | |
| | 2010 | | | | | | | | | 0 | | | 1 | 2 | | | | 3 | | | |
| Dried vegetables, cut into pieces / slices, crushed / powder, etc. | 2006 | | | | | | | 2 | | 0 | | | 14 | 8 | | 3 | | 1 | | 4 | 1 |
| | 2007 | | | | | | | 0 | | 2 | | | 23 | 9 | | 3 | | 1 | | 3 | 0 |
| | 2008 | | | | | | | 0 | | 2 | | | 7 | 20 | | 2 | 1 | 2 | | 5 | 1 |
| | 2009 | | | | | | | 0 | | | | 100 | 15 | 17 | | 3 | | 1 | | 4 | 1 |
| | 2010 | | | | | | | 0 | | 2 | | | 18 | 34 | 19 | 6 | | 1 | | 3 | 1 |
| Leguminous vegetables, dried or shelled, peeled or broken | 2006 | | | | | | | 4 | 14 | 100 | 100 | | 81 | 0 | | 1 | | 2 | | 1 | 4 |
| | 2007 | | | 100 | | | | 84 | 99 | 98 | | | 71 | 0 | | 0 | | 2 | | 0 | 4 |
| | 2008 | | | 100 | | | | 72 | 100 | 98 | | | 89 | 1 | | 0 | | 3 | 18 | 0 | 4 |
| | 2009 | | | 100 | | | | 8 | 72 | 100 | 100 | | 77 | 0 | | 0 | | 6 | | 1 | 2 |
| | 2010 | | | 100 | | | | 62 | 53 | 100 | 97 | | 78 | 1 | 0 | 1 | 3 | 3 | 2 | 1 | 2 |
| Manioc, sweet potatoes and similar roots and tubers, fresh or dried, | 2006 | | 1 | | | | | 31 | | | | | 1 | 0 | | | | 0 | | 0 | 3 |
| | 2007 | | | | | | | 2 | | | | | 0 | 0 | | | | 1 | | 0 | 1 |
| | 2008 | | | | | | | 6 | | | | | 0 | 0 | | 0 | 1 | 0 | 1 | 0 | 2 |
| | 2009 | | | | | | | 28 | | | | | 1 | 0 | | 1 | | 0 | | 0 | 1 |
| | 2010 | | | | | | | 44 | | | | | 1 | 0 | | 1 | | 0 | | 0 | 1 |

Observing the table 2 it is shown that Portugal export for Angola, Cape Verde, Guinea-Bissau, Mozambique, Sao Tome and Principe Italy several horticultural products, but principally leguminous vegetables, dried or shelled, peeled or broken.

The exports for Morocco are insignificant. To Brazil and Canada, the Portuguese export different vegetables, without some evident predominance.

For the United States, Switzerland, China, Austria, Belgium, Denmark, Estonia, France and Luxembourg export vegetables, uncooked or cooked by steaming, frozen.

To Germany and Ireland Portugal export various horticultural products, but mainly dried vegetables, cut into pieces/slices, crushed/powder, etc. To Spain export tomatoes, to Holland and United Kingdom export other vegetables. Portugal export the same horticultural products that import, again is important find a coherent policy to the horticultural sector.

**Table 2. Horticultural products, in different forms, export percentage relatively to the total of each country**

| | year | Angola | Cape Verde | Guinea-Bissau | Mozambique | Sao Tome and Principe | Morocco | Brazil | Canada | United States of America | Switzerland | China | Germany | Austria | Belgium | Denmark | Spain | Estonia | France | Holland |
|---|---|---|---|---|---|---|---|---|---|---|---|---|---|---|---|---|---|---|---|---|
| Potatoes, fresh or chilled | 2006 | 3 | 31 | 2 | | 6 | 100 | | | 2 | | 94 | 15 | | 1 | | 13 | | 2 | 27 |
| | 2007 | 2 | 29 | | | 4 | | | | | | | 18 | | 1 | | 15 | | 6 | 20 |
| | 2008 | 1 | 22 | 28 | | 4 | | | | | | | 23 | | 0 | | 8 | | 6 | 34 |
| | 2009 | 2 | 17 | 35 | | 4 | | | | 9 | | | 21 | | 2 | 0 | 9 | | 8 | 28 |
| | 2010 | 3 | 20 | 42 | 4 | 6 | | | | 9 | | | 56 | | 3 | | 7 | | 3 | 27 |
| Tomatoes, fresh or chilled | 2006 | 0 | 2 | | | 0 | | | | | 3 | | 0 | 0 | | | 27 | | 2 | 0 |
| | 2007 | 0 | 2 | | | | | | | | 2 | | 0 | | | | 33 | | 6 | |
| | 2008 | 0 | 2 | 1 | | 1 | | | | | | | | | | | 36 | | 4 | 0 |
| | 2009 | 0 | 2 | 1 | | | | | | | | | 0 | | | | 40 | | 4 | |
| | 2010 | 0 | 3 | 0 | | | | | | | | | 0 | | | | 28 | | 8 | |
| Onions, shallots and other prod vegetables, fresh or chilled | 2006 | 1 | 7 | | 0 | 1 | | | | 0 | | 6 | 3 | | | | 7 | | 8 | 2 |
| | 2007 | 1 | 6 | | | 1 | | | 0 | 0 | | | 1 | | 0 | | 3 | | 2 | 3 |
| | 2008 | 0 | 3 | 5 | | 1 | | | | 0 | | | 1 | | | | 3 | | 2 | 5 |
| | 2009 | 0 | 3 | 5 | 0 | 5 | | | | 0 | 0 | | 1 | | 0 | | 2 | | 4 | 4 |
| | 2010 | 1 | 11 | 9 | 0 | 7 | | | | 0 | 0 | | 2 | | 0 | 1 | 3 | | 5 | 3 |
| Cabbages, cauliflower, cabbage or kale kohlrabi, etc., Fresh or chilled | 2006 | 1 | 2 | | | | | | | | 18 | | 20 | | 0 | | 4 | | 5 | 9 |
| | 2007 | 1 | 2 | | | | | | | | 28 | | 16 | | | 3 | 2 | | 5 | 10 |
| | 2008 | 1 | 2 | 3 | | | | | 6 | | 18 | | 12 | | 0 | 1 | 2 | | 3 | 11 |
| | 2009 | 1 | 2 | 3 | 0 | 0 | | | 2 | | 7 | | 16 | | 0 | 8 | 2 | | 3 | 17 |
| | 2010 | 1 | 2 | 3 | | 0 | | | 5 | | 3 | | 8 | | 0 | 2 | 2 | | 3 | 13 |
| Lettuce and chicory, fresh or chilled | 2006 | 0 | 0 | | | | | | | | | | | | 1 | | 6 | | 14 | 1 |
| | 2007 | 0 | 0 | | | | | | | | | | 0 | | 1 | | 6 | | 10 | 0 |
| | 2008 | 0 | 0 | | | | | | | | | | 6 | | 2 | | 8 | | 6 | 2 |
| | 2009 | 0 | 0 | | | | | | | | | | 0 | | 0 | 2 | 8 | | 5 | 1 |
| | 2010 | 0 | 0 | | | | | | | | | | 0 | | | 1 | 10 | | 3 | 0 |
| Carrots, turnips, salad beetroot, radishes, etc., fresh or chilled | 2006 | 0 | 6 | | | 0 | | | | | 0 | | 10 | | 1 | | 0 | | 6 | 1 |
| | 2007 | 0 | 8 | | | | | | | | | | 12 | | 1 | | 1 | | 4 | 0 |
| | 2008 | 0 | 9 | 3 | | | | | | | | | 10 | | 2 | | 1 | | 5 | 0 |
| | 2009 | 0 | 8 | 1 | | | | | | | | | 13 | | 2 | | 1 | | 10 | 2 |
| | 2010 | 0 | 8 | 1 | | 0 | | | | | 0 | | 20 | | 0 | | 1 | | 4 | 2 |
| Cucumbers and gherkins (gherkins), fresh or chilled | 2006 | | 0 | | | | | | | | | | | | | | 0 | | 0 | |
| | 2007 | | 0 | | | | | | | | | | 0 | | | | 1 | | 0 | |
| | 2008 | 0 | 0 | | | | | | | | | | | | | | 0 | | 0 | |
| | 2009 | 0 | 0 | | | | | | | | | | 0 | | | | 0 | | 0 | |
| | 2010 | 0 | 0 | | | | | | | | | | 0 | | | | 0 | | 0 | |
| Leguminous vegetables, shelled or unshelled, fresh or chilled | 2006 | 1 | 2 | 0 | 0 | 1 | | | 2 | 2 | | | 0 | | 0 | | 0 | | 0 | |
| | 2007 | 1 | 0 | | 1 | 0 | | | | 0 | 0 | | 0 | | | | 1 | | 0 | |
| | 2008 | 0 | 0 | | 1 | 0 | | | | 1 | 0 | | | | 0 | | 1 | | 0 | 0 |
| | 2009 | 0 | 0 | 1 | 8 | 0 | | 0 | | | | | | | 0 | | 1 | | 0 | 0 |
| | 2010 | 1 | 0 | 0 | 0 | 0 | | | | | 0 | | 0 | | 0 | | 0 | | | |
| Other vegetables, fresh or chilled | 2006 | 1 | 2 | 2 | 1 | 1 | | 55 | 10 | 32 | 27 | | 1 | 0 | 2 | 1 | 11 | | 13 | 43 |
| | 2007 | 1 | 2 | 0 | 0 | 0 | | 4 | 83 | 39 | 15 | 0 | 1 | | 2 | 4 | 9 | | 12 | 42 |
| | 2008 | 1 | 4 | 3 | 1 | 1 | | 75 | 14 | 24 | | 0 | 0 | | 2 | 4 | 13 | | 19 | 37 |
| | 2009 | 2 | 20 | 1 | 3 | 0 | | 0 | 4 | 2 | 2 | | 2 | 0 | 2 | 19 | 12 | | 17 | 39 |
| | 2010 | 2 | 10 | 2 | 3 | 0 | | 72 | 89 | 13 | 20 | | 4 | 0 | 2 | 19 | 19 | | 30 | 47 |
| Vegetables, uncooked or cooked by steaming, frozen | 2006 | 23 | 2 | | 0 | 2 | | 16 | | 49 | 31 | | 5 | 100 | 93 | 94 | 12 | | 48 | 7 |
| | 2007 | 19 | 6 | 9 | 4 | 0 | | 12 | | 55 | 20 | | 9 | 100 | 94 | 85 | 16 | | 51 | 3 |
| | 2008 | 18 | 11 | 0 | 50 | 2 | | 49 | 6 | 81 | 37 | 99 | 14 | 100 | 93 | 80 | 12 | 100 | 52 | 3 |
| | 2009 | 20 | 7 | 2 | 22 | 1 | | 81 | 89 | 95 | 58 | 100 | 11 | 100 | 93 | 65 | 6 | 100 | 45 | 3 |
| | 2010 | 12 | 8 | 3 | 27 | 2 | 100 | 23 | | 76 | 46 | 100 | 3 | 100 | 93 | 74 | 6 | 100 | 38 | 2 |
| Production vegetables preserved, but not for food in this state | 2006 | 1 | 0 | | 0 | 0 | | 18 | 14 | 4 | | | 0 | | | | 2 | | 0 | |
| | 2007 | 0 | 0 | 1 | | 1 | | 63 | | 4 | | | | | | | 1 | | 0 | |
| | 2008 | 0 | 1 | 0 | | 0 | | 12 | | 1 | | | | | | | 1 | | 0 | |
| | 2009 | 0 | 1 | | 4 | 0 | | 3 | | | | | 0 | | | 0 | 1 | | 0 | |
| | 2010 | 0 | 2 | 6 | 4 | 0 | | 5 | | | | | | | | 0 | 0 | | 0 | |
| Dried vegetables, cut into pieces / slices, crushed / powder, etc. | 2006 | 2 | 0 | 0 | 1 | 1 | | 27 | 0 | | 0 | | 44 | | 1 | 5 | 1 | | 2 | 10 |
| | 2007 | 0 | 1 | | 2 | 0 | | 21 | 0 | 0 | 0 | 100 | 42 | | 1 | 8 | 0 | | 2 | 21 |
| | 2008 | 1 | 2 | 4 | 0 | 0 | | 5 | | | 0 | 1 | 33 | | 1 | 15 | 0 | | 2 | 9 |
| | 2009 | 2 | 1 | 6 | 0 | 0 | | 16 | | 0 | 0 | | 35 | | 0 | 5 | 0 | | 1 | 5 |
| | 2010 | 2 | 5 | 1 | 1 | 2 | | 0 | | 5 | 0 | | 8 | | 1 | 3 | 1 | | 3 | 5 |
| Leguminous vegetables, dried or shelled, peeled or broken | 2006 | 68 | 41 | 95 | 98 | 88 | | | 34 | 3 | | 19 | 0 | | 0 | | 16 | | 1 | 0 |
| | 2007 | 75 | 39 | 90 | 93 | 94 | | | | 0 | | 34 | 0 | 0 | 0 | | 12 | | 1 | 0 |
| | 2008 | 77 | 39 | 54 | 45 | 91 | | 34 | | 0 | | 20 | 0 | 0 | 0 | | 14 | | 1 | 0 |
| | 2009 | 71 | 38 | 44 | 62 | 90 | | | 1 | 1 | | 24 | 0 | 0 | 0 | | 17 | | 1 | 0 |
| | 2010 | 77 | 28 | 32 | 61 | 82 | | 6 | 0 | 0 | | 20 | 0 | 0 | 0 | | 22 | | 2 | 1 |
| Manioc, sweet potatoes and similar roots and tubers, fresh or dried | 2006 | | 4 | | | | | | 24 | 9 | | | | | | | 0 | | 0 | |
| | 2007 | | 4 | | | | | | 17 | 2 | | | | | | | 0 | | 0 | |
| | 2008 | 0 | 4 | | 2 | | | | 13 | 2 | | | 0 | | | | 0 | | 0 | |
| | 2009 | | 2 | | | | | | 4 | 3 | | | | | | | 0 | | 0 | 0 |
| | 2010 | 0 | 1 | | | | | | | 5 | 0 | | 0 | | | | 0 | | 0 | 0 |

Portugal imports the majority of the vegetables from Spain (table 3). Import a significant part of potatoes from France, vegetables cooked or uncooked from Belgium, dried vegetables from Germany and France, leguminous from China and Canada, and manioc, sweet potatoes and similar roots and tubers, fresh or dried, etc, from Holland.

**Table 3. Horticultural products, in different forms, import percentage relatively to the total of each year**

| | year | Angola | Cape Verde | Guinea-Bissau | Mozambique | Sao Tome and Principe | Morocco | Brazil | Canada | United States of America | Switzerland | China | Germany | Austria | Belgium | Denmark | Spain | Estonia | France | Holland |
|---|---|---|---|---|---|---|---|---|---|---|---|---|---|---|---|---|---|---|---|---|
| Potatoes, fresh or chilled | 2006 | | | | | | 0 | | 0 | | | | 3 | | 1 | 2 | 33 | 0 | 41 | 17 |
| | 2007 | | | | | | 0 | | 0 | | | | 3 | | 1 | 4 | 31 | | 39 | 19 |
| | 2008 | | | | | | | | 0 | | | | 4 | | 2 | 2 | 40 | | 32 | 18 |
| | 2009 | | | | | | | | | | | | 5 | | 2 | 3 | 32 | | 38 | 19 |
| | 2010 | | | | | | 0 | | 0 | | | | 5 | | 1 | 1 | 38 | 0 | 38 | 15 |
| Tomatoes, fresh or chilled | 2006 | | | | | | 0 | 0 | | | | | 15 | | 1 | | 80 | | 2 | 1 |
| | 2007 | | | | | | | | | | | | 11 | | 0 | | 87 | | 2 | 1 |
| | 2008 | | | | | | | | | | | | 0 | | 0 | | 93 | 0 | 6 | 1 |
| | 2009 | | | | | | | | | | | | 0 | | 0 | | 96 | 0 | 2 | 1 |
| | 2010 | | | | | | 0 | | | | | | 0 | | 0 | | 98 | 0 | 1 | 0 |
| Onions, shallots and other prod vegetables, fresh or chilled | 2006 | | | | | | | | | | | 0 | 17 | | 1 | 0 | 65 | 0 | 11 | 3 |
| | 2007 | | | | | | 0 | 0 | | | | 0 | 14 | | 0 | 0 | 66 | | 15 | 4 |
| | 2008 | | | | | | | | | | | 0 | 4 | | 0 | | 80 | 0 | 14 | 1 |
| | 2009 | | | | | | | | | | | | 1 | | 0 | | 93 | 0 | 5 | 1 |
| | 2010 | | | | | | | | | | 0 | 0 | 0 | | 1 | | 88 | 0 | 8 | 3 |
| Cabbages, cauliflower, cabbage or kale kohlrabi, etc., Fresh or chilled | 2006 | | | | | | | | | | | 0 | 15 | | 1 | | 70 | | 9 | 4 |
| | 2007 | | | | | | | | | | | | 17 | | 2 | | 62 | | 16 | 3 |
| | 2008 | | | | | | | | | | | 0 | 1 | | 1 | | 82 | 0 | 6 | 10 |
| | 2009 | | | | | | | | | | | | 1 | | 1 | | 85 | 0 | 8 | 6 |
| | 2010 | | | | | | | | | | | | 0 | | 1 | | 91 | 0 | 4 | 4 |
| Lettuce and chicory, fresh or chilled | 2006 | | | | | | | | | | | | 1 | | 6 | | 81 | | 11 | 2 |
| | 2007 | | | | | | | | | | | | 1 | | 13 | | 81 | | 2 | 2 |
| | 2008 | | | | | | | | | | | | 1 | | 10 | | 83 | 0 | 3 | 2 |
| | 2009 | | | | | | | | | | | | 0 | | 7 | | 79 | | 10 | 2 |
| | 2010 | | | | | | | | | | | | 1 | | 3 | | 76 | 0 | 13 | 2 |
| Carrots, turnips, salad beetroot, radishes, etc., fresh or chilled | 2006 | | | | | | | | | | | | 13 | | 4 | | 50 | | 31 | 1 |
| | 2007 | | | | | | | | | | | 0 | 16 | | 6 | | 59 | | 19 | 1 |
| | 2008 | | | | | | | | | | | 0 | 1 | | 1 | | 84 | 0 | 13 | 1 |
| | 2009 | | | | | 0 | | | | | | | 0 | | 3 | | 83 | 0 | 12 | 1 |
| | 2010 | | | | | | | | | | | | 0 | | 6 | | 73 | 0 | 19 | 1 |
| Cucumbers and gherkins (gherkins), fresh or chilled | 2006 | | | | 0 | | | | | | | | 0 | | 0 | | 97 | | 0 | 3 |
| | 2007 | | | | 0 | | | | | | | | 1 | | 1 | | 96 | | 1 | 1 |
| | 2008 | | | | 0 | | | | | | | | 0 | | 0 | | 99 | 0 | 0 | 1 |
| | 2009 | | | | 1 | | | | | | | | 0 | | 0 | | 98 | 0 | | 0 |
| | 2010 | | | | 0 | | | | | | | | 0 | | 0 | | 99 | | 0 | 0 |
| Leguminous vegetables, shelled or unshelled, fresh or chilled | 2006 | | | | 1 | | | | | | | | 3 | | 0 | | 91 | | 4 | 0 |
| | 2007 | | | | 1 | | | 0 | | | | | 2 | | 0 | | 92 | | 5 | 0 |
| | 2008 | | | | 1 | | | 0 | | | | | 0 | | 0 | | 98 | 0 | 0 | 0 |
| | 2009 | | | | 2 | | | | | | | | 0 | | 0 | | 98 | 0 | 0 | 0 |
| | 2010 | | | | 2 | | | | | | | | 0 | | 1 | | 96 | 0 | 0 | 0 |
| Other vegetables, fresh or chilled | 2006 | | 0 | | 0 | 1 | | | | | | 0 | 5 | 1 | 0 | | 84 | 0 | 2 | 6 |
| | 2007 | | 0 | | 0 | 1 | | | | | | | 4 | 0 | 1 | | 87 | 0 | 2 | 5 |
| | 2008 | | 0 | | 0 | 0 | | | | | | 0 | 1 | 0 | 1 | | 88 | 0 | 4 | 5 |
| | 2009 | | 0 | 0 | 0 | 0 | | | | | | 0 | 2 | 0 | 2 | | 86 | 0 | 3 | 6 |
| | 2010 | | 0 | | 0 | 0 | | | 0 | | | 0 | 1 | 0 | 1 | 0 | 89 | 0 | 2 | 5 |
| Vegetables, uncooked or cooked by steaming, frozen | 2006 | | | | | | | | | | | 0 | 2 | 0 | 27 | | 43 | 0 | 15 | 5 |
| | 2007 | | | | | | | | | | | 0 | 2 | 0 | 24 | | 46 | | 15 | 5 |
| | 2008 | | | | | | | | | | | 0 | 1 | 0 | 23 | | 47 | 0 | 13 | 7 |
| | 2009 | | | | 0 | | | | | | | 0 | 4 | 0 | 21 | | 50 | 0 | 11 | 7 |
| | 2010 | | | | | | | | | | 0 | 0 | 3 | 0 | 23 | | 49 | 0 | 13 | 4 |
| Production vegetables preserved, but not for food in this state | 2006 | | | | | | | | | | | 0 | 2 | | 0 | 0 | 96 | | 0 | 1 |
| | 2007 | | | | | | | | | 0 | | 1 | 3 | | | | 95 | | 0 | 0 |
| | 2008 | | | | | | | | | 0 | | 2 | 2 | | 0 | | 96 | | 0 | 0 |
| | 2009 | | | | | | | | | 0 | | 2 | 3 | | | 0 | 95 | 0 | | |
| | 2010 | | | | | | | | | 0 | | 1 | 3 | | | | 96 | | | |
| Dried vegetables, cut into pieces / slices, crushed / powder, etc. | 2006 | | | | | | | 0 | | 0 | | 10 | 23 | | 6 | | 21 | | 36 | 3 |
| | 2007 | | | | | | | 0 | | 1 | | 14 | 29 | | 6 | | 23 | | 25 | 1 |
| | 2008 | | | | | | | 0 | | 1 | | 9 | 20 | | 3 | 0 | 38 | | 27 | 2 |
| | 2009 | | | | | | | | | 0 | 0 | 12 | 21 | | 5 | | 35 | | 24 | 2 |
| | 2010 | | | | | | | 0 | | 0 | | 14 | 34 | 0 | 9 | | 25 | | 17 | 1 |
| Leguminous vegetables, dried or shelled, peeled or broken | 2006 | | | | | | | 0 | 40 | 8 | | 20 | 0 | | 0 | | 19 | | 3 | 5 |
| | 2007 | | | 0 | | | | 6 | 47 | 9 | | 12 | 0 | | 0 | | 19 | | 1 | 4 |
| | 2008 | | | 0 | | | | 1 | 35 | 7 | | 33 | 0 | | 0 | | 17 | 0 | 0 | 4 |
| | 2009 | | | 1 | | | | 1 | 30 | 3 | | 16 | 0 | | 0 | | 44 | | 1 | 2 |
| | 2010 | | | 0 | | | | 1 | 37 | 8 | | 24 | 0 | 0 | 0 | 0 | 21 | 0 | 2 | 2 |
| Manioc, sweet potatoes and similar roots and tubers, fresh or dried, | 2006 | | 0 | | | | | 10 | | | | | 3 | 0 | | | 22 | | 15 | 50 |
| | 2007 | | | | | | | 2 | | | | | 1 | 0 | | | 65 | | 9 | 21 |
| | 2008 | | | | | | | 2 | | | | | 2 | 0 | 0 | 3 | 50 | 0 | 3 | 37 |
| | 2009 | | | | | | | 9 | | | | | 3 | 0 | 11 | | 41 | | 1 | 26 |
| | 2010 | | | | | | | 17 | | | | | 4 | 0 | 9 | | 41 | | 2 | 18 |

The exports of the Portuguese vegetables are principally, also, to Spain (table 4). Export, yet, an important part of potatoes and tomatoes to the United Kingdom, onions to France, cabbages, cauliflower, cabbage or kale kohlrabi, etc, fresh or chilled to Germany, France, Holland and United Kingdom, lettuce and chicory, fresh or chilled and carrots, turnips, salad beetroot, radishes, etc, fresh or chilled to France and United Kingdom, other vegetables to United Kingdom, vegetables cooked or uncooked to Belgium and France and dried vegetables to the United Kingdom.

**Table 4. Horticultural products, in different forms, export percentage relatively to the total of each year**

| | year | Angola | Cape Verde | Guinea-Bissau | Mozambique | Sao Tome and Principe | Morocco | Brazil | Canada | United States of America | Switzerland | China | Germany | Austria | Belgium | Denmark | Spain | Estonia | France | Holland |
|---|---|---|---|---|---|---|---|---|---|---|---|---|---|---|---|---|---|---|---|---|
| Potatoes, fresh or chilled | 2006 | 0 | 5 | 0 | | 0 | 0 | | | | 0 | 0 | 10 | | 1 | | 31 | | 2 | 15 |
| | 2007 | 0 | 4 | | | 0 | | | | | | | 8 | | 1 | | 38 | | 7 | 9 |
| | 2008 | 0 | 3 | 0 | | 0 | | | | | | | 16 | | 0 | | 26 | | 7 | 19 |
| | 2009 | 1 | 3 | 0 | | 0 | | | | | 0 | | 19 | | 2 | 0 | 35 | | 12 | 17 |
| | 2010 | 1 | 4 | 0 | 0 | 0 | | | | | 1 | | 43 | | 3 | | 25 | | 4 | 18 |
| Tomatoes, fresh or chilled | 2006 | 0 | 0 | | | 0 | | | | | 0 | | 0 | 0 | | | 65 | | 3 | 0 |
| | 2007 | 0 | 0 | | | | | | | | 0 | | 0 | | | | 71 | | 6 | |
| | 2008 | 0 | 0 | 0 | | 0 | | | | | | | | | | | 80 | | 4 | 0 |
| | 2009 | 0 | 0 | 0 | | | | | | | | | 0 | | | | 78 | | 3 | |
| | 2010 | 0 | 0 | 0 | | | | | | | | | 0 | | | | 67 | | 6 | |
| Onions, shallots and other prod vegetables, fresh or chilled | 2006 | 0 | 3 | | 0 | 0 | | | | 0 | | 0 | 5 | | | | 45 | | 32 | 3 |
| | 2007 | 1 | 5 | | | 0 | | | | 0 | 0 | | 3 | | 0 | | 55 | | 18 | 8 |
| | 2008 | 1 | 3 | 0 | | 0 | | | | | 0 | | 3 | | | | 52 | | 16 | 13 |
| | 2009 | 0 | 3 | 0 | 0 | 1 | | | | 0 | 0 | | 5 | | 0 | | 42 | | 33 | 14 |
| | 2010 | 1 | 9 | 0 | 0 | 1 | | | | 0 | 0 | | 6 | | 0 | 0 | 43 | | 23 | 7 |
| Cabbages, cauliflower, cabbage or kale kohlrabi, etc... Fresh or chilled | 2006 | 0 | 1 | | | | | | | | 2 | | 27 | | 1 | | 17 | | 16 | 11 |
| | 2007 | 1 | 1 | | | | | | | | 3 | | 19 | | | 1 | 18 | | 16 | 14 |
| | 2008 | 1 | 1 | 0 | | | | 0 | | | 2 | | 23 | | 0 | | 19 | | 12 | 17 |
| | 2009 | 1 | 1 | 0 | 0 | 0 | | 0 | | | 1 | | 30 | | 0 | 1 | 13 | | 9 | 21 |
| | 2010 | 1 | 1 | 0 | | 0 | | 0 | | | 0 | | 16 | | 0 | 0 | 21 | | 9 | 22 |
| Lettuce and chicory, fresh or chilled | 2006 | 0 | 0 | | | | | | | | | | | | 2 | | 19 | | 32 | 1 |
| | 2007 | 0 | 0 | | | | | | | | | | 0 | | 1 | | 29 | | 23 | 0 |
| | 2008 | 0 | 0 | | | | | | | | | | 7 | | 3 | | 48 | | 14 | 2 |
| | 2009 | 0 | 0 | | | | | | | | | | 1 | | 0 | 0 | 53 | | 14 | 1 |
| | 2010 | 0 | 0 | | | | | | | | | | 0 | | | 0 | 73 | | 7 | 0 |
| Carrots, turnips, salad beetroot, radishes, etc., fresh or chilled | 2006 | 0 | 3 | | | 0 | | | | | 0 | | 19 | | 3 | | 3 | | 26 | 1 |
| | 2007 | 0 | 4 | | | | | | | | | | 16 | | 3 | | 9 | | 15 | 0 |
| | 2008 | 0 | 3 | 0 | | | | | | | | | 17 | | 4 | | 5 | | 17 | 0 |
| | 2009 | 0 | 3 | 0 | | | | | | | | | 24 | | 5 | | 7 | | 29 | 2 |
| | 2010 | 0 | 6 | 0 | | 0 | | | | | 0 | | 46 | | 1 | | 9 | | 15 | 4 |
| Cucumbers and gherkins (gherkins), fresh or chilled | 2006 | | 7 | | | | | | | | | | | | | | 71 | | 21 | |
| | 2007 | | 1 | | | | | | | | | | 0 | | | | 95 | | 4 | |
| | 2008 | 0 | 3 | | | | | | | | | | | | | | 77 | | 8 | |
| | 2009 | 5 | 2 | | | | | | | | | | 0 | | | | 87 | | 3 | |
| | 2010 | 1 | 3 | | | | | | | | | | 0 | | | | 84 | | 4 | |
| Leguminous vegetables, shelled or unshelled, fresh or chilled | 2006 | 4 | 11 | 0 | 0 | 1 | | | 0 | 1 | | | 0 | | 0 | | 28 | | 9 | |
| | 2007 | 4 | 1 | 0 | 0 | 0 | | | 0 | | 0 | | 0 | | 0 | | 58 | | 8 | |
| | 2008 | 1 | 1 | 0 | 0 | 0 | | | 1 | | 0 | | | | | | 94 | | 0 | 0 |
| | 2009 | 4 | 1 | 0 | 2 | 0 | | 0 | | | | | | | 0 | | 85 | | 4 | 0 |
| | 2010 | 10 | 5 | 0 | 0 | 0 | | | | | 1 | | 0 | | 0 | | 79 | | 1 | |
| Other vegetables, fresh or chilled | 2006 | 0 | 0 | 0 | 0 | 0 | | 0 | 0 | 0 | 1 | | 0 | 1 | 0 | 1 | 18 | | 13 | 16 |
| | 2007 | 0 | 0 | 0 | 0 | 0 | | 0 | 0 | 1 | 0 | 0 | 0 | | 1 | 0 | 19 | | 12 | 16 |
| | 2008 | 0 | 0 | 0 | 0 | 0 | | | 0 | 0 | 1 | | 0 | 0 | 2 | 0 | 31 | | 17 | 15 |
| | 2009 | 0 | 2 | 0 | 0 | 0 | | 0 | 0 | 0 | 0 | | 1 | 0 | 2 | 1 | 29 | | 16 | 15 |
| | 2010 | 0 | 1 | 0 | 0 | 0 | | 0 | 0 | 0 | 1 | | 1 | 0 | 1 | 1 | 33 | | 17 | 15 |
| Vegetables, uncooked or cooked by steaming, frozen | 2006 | 2 | 0 | | 0 | 0 | | | 0 | 0 | 1 | | 1 | 1 | 38 | 5 | 13 | | 31 | 2 |
| | 2007 | 2 | 0 | 0 | 0 | 0 | | 0 | | 1 | 0 | | 2 | 1 | 37 | 3 | 20 | | 29 | 1 |
| | 2008 | 2 | 1 | 0 | 0 | 0 | | 0 | 0 | 1 | 1 | 0 | 4 | 1 | 39 | 1 | 16 | 0 | 28 | 1 |
| | 2009 | 2 | 1 | 0 | 0 | 0 | | 0 | 0 | 1 | 1 | 0 | 4 | 1 | 43 | 2 | 10 | 0 | 29 | 1 |
| | 2010 | 2 | 1 | 0 | 0 | 0 | 0 | 0 | | 0 | 1 | 0 | 1 | 3 | 46 | 3 | 12 | 0 | 22 | 1 |
| Production vegetables preserved, but not for food in this state | 2006 | 3 | 0 | | 0 | 0 | | 1 | 1 | 1 | | | 0 | | | | 89 | | 5 | |
| | 2007 | 0 | 1 | 0 | | 0 | | 1 | | 2 | | | | | | | 89 | | 3 | |
| | 2008 | 3 | 7 | 0 | | 0 | | 1 | | 1 | | | | | | | 82 | | 4 | |
| | 2009 | 1 | 2 | | 0 | 0 | | 0 | | | | | 0 | | | | 93 | | 2 | |
| | 2010 | 3 | 28 | 1 | 1 | 0 | | 1 | | | | | | | 0 | | 65 | | 1 | |
| Dried vegetables, cut into pieces / slices, crushed / powder, etc. | 2006 | 1 | 0 | 0 | 0 | 0 | | 0 | 0 | | 0 | | 49 | | 2 | 1 | 6 | | 6 | 10 |
| | 2007 | 0 | 0 | 0 | 0 | 0 | | 0 | 0 | 0 | 0 | 1 | 42 | | 2 | 1 | 1 | | 6 | 24 |
| | 2008 | 1 | 1 | 0 | 0 | 0 | | 0 | | | 0 | 0 | 51 | | 1 | 1 | 1 | | 6 | 12 |
| | 2009 | 1 | 0 | 0 | 0 | 0 | | 0 | | 0 | 0 | | 61 | | 0 | 1 | 1 | | 4 | 5 |
| | 2010 | 2 | 3 | 0 | 0 | 0 | | 0 | | | 0 | | 19 | | 3 | 1 | 11 | | 11 | 10 |
| Leguminous vegetables, dried or shelled, peeled or broken | 2006 | 16 | 10 | 0 | 1 | 3 | | 0 | 0 | | 2 | | 0 | | 0 | | 54 | | 1 | 0 |
| | 2007 | 24 | 8 | 0 | 1 | 4 | | | 0 | | 2 | 0 | 0 | | | | 48 | | 2 | 0 |
| | 2008 | 25 | 7 | 0 | 0 | 3 | | 0 | 0 | | 1 | | 0 | 0 | 0 | | 50 | | 1 | 0 |
| | 2009 | 17 | 6 | 0 | 0 | 3 | | 0 | 0 | | 1 | | 0 | 0 | 0 | | 63 | | 1 | 0 |
| | 2010 | 21 | 5 | 0 | 0 | 2 | | 0 | 0 | | 1 | | 0 | 0 | 0 | | 61 | | 1 | 0 |
| Manioc, sweet potatoes and similar roots and tubers, fresh or dried, | 2006 | | 48 | | | | | 7 | 8 | | | | | | | | 16 | | 15 | |
| | 2007 | | 50 | | | | | 1 | 4 | | | | | | | | 33 | | 11 | |
| | 2008 | 0 | 37 | | 1 | | | 0 | 6 | | | | 7 | | | | 49 | | | |
| | 2009 | | 21 | | | | | 1 | 5 | | | | | | | | 62 | | 0 | 1 |
| | 2010 | 1 | 9 | | | | | 2 | 0 | | | | 2 | | | | 55 | | 4 | |

Portugal import and export horticultural products principally to the European countries, what is comprehensive, taking into account the costs of transport vegetables and what is said by the new economic geography, the cost of transport are important in the economic dynamic.

## 4. Estimations results for the neoclassical model with panel data

The results presented in the tables 5 to 10 are obtained with panel data estimations, with different method of estimations and taking into account the neoclassical theory based in the Solow (1956) and Islam (1995) developments, described in various works.

The results shown in the following six tables evidence convergence of the data in all estimations, because the results obtained from the different estimations are statistically significant either in the coefficients and in the global model, taking into account the several statistics tests.

**Table 5. Results from the absolute convergence model for all vegetables import (absolute values)**

|  | Const.[1] | Coef.[2] | F/Wald(mod.)[3] | F(Fe_OLS)[4] | Corr(u_i)[5] | F(Re_OLS)[6] | Hausman[7] | $R^2$[8] | N.O.[9] | N.I.[10] |
|---|---|---|---|---|---|---|---|---|---|---|
| FE[11] | 11.188* (18.120) | -0.943* (-18.240) | 332.560* | 3.120* | -0.929 | ------- | ------- | 0.512 | 449 | ------- |
| RE[12] | 1.571* (5.710) | -0.135* (-6.020) | 36.200* | ------- | ------- | 9.330* | 300.850* | 0.512 | 449 | ------- |
| OLS | ------- | ------- | ------- | ------- | ------- | ------- | ------- | ------- | ------- | ------- |
| DPD[13] | 18.266* (19.970) | -1.517* (-20.140) | 507.640* | ------- | ------- | ------- | ------- | ------- | 200 | 5 |

Note: 1, Constant; 2, Coefficient; 3, Test F for fixed effects model and test Wald for random effects and dynamic panel data models; 4, Test F for fixed effects or OLS (Ho is OLS); 5, Correlation between errors and regressors in fixed effects; 6, Test F for random effects or OLS (Ho is OLS); 7, Hausman test (Ho is GLS); 8, R square; 9, Number of observations; 10, Number of instruments;, 11, Fixed effects model; 12, Random effects model; 13, Dynamic panel data model; *, Statically significant at 5%.

The estimations results presented in these tables show that there are strong signs of convergence between the several horticultural products, imported and exported by Portugal from and for, respectively, different countries, from 2006 to 2010. All estimations and all statistically tests show which these data are influenced no by random effects, but by fixed effects.

**Table 6. Results from the absolute convergence model for all vegetables export (absolute values)**

|  | Const.[1] | Coef.[2] | F/Wald(mod.)[3] | F(Fe_OLS)[4] | Corr(u_i)[5] | F(Re_OLS)[6] | Hausman[7] | $R^2$[8] | N.O.[9] | N.I.[10] |
|---|---|---|---|---|---|---|---|---|---|---|
| FE[11] | 11.760* (23.240) | -1.077* (-23.100) | 533.770* | 3.590* | -0.939 | ------- | ------- | 0.544 | 641 | ------- |
| RE[12] | 2.518* (9.570) | -0.226* (-9.570) | 91.640* | ------- | ------- | 16.210* | 448.040* | 0.544 | 641 | ------- |
| OLS | ------- | ------- | ------- | ------- | ------- | ------- | ------- | ------- | ------- | ------- |
| DPD[13] | 19.201* (18.480) | -1.712* (-18.350) | 543.570* | ------- | ------- | ------- | ------- | ------- | 277 | 5 |

**Table 7. Results from the absolute convergence model for all vegetables import (percentage values relatively to the total of each country)**

|  | Const.[1] | Coef.[2] | F/Wald(mod.)[3] | F(Fe_OLS)[4] | Corr(u_i)[5] | F(Re_OLS)[6] | Hausman[7] | $R^2$[8] | N.O.[9] | N.I.[10] |
|---|---|---|---|---|---|---|---|---|---|---|
| FE[11] | 0.824* (13.150) | -1.049* (-19.920) | 396.630* | 3.580* | -0.918 | ------- | ------- | 0.556 | 449 | ------- |
| RE[12] | 0.120 (1.610) | -0.195* (-7.070) | 49.970* | ------- | ------- | 9.330* | 361.970* | 0.556 | 449 | ------- |
| OLS | ------- | ------- | ------- | ------- | ------- | ------- | ------- | ------- | ------- | ------- |
| DPD[13] | 1.445* (16.310) | -1.682* (-19.630) | 488.880* | ------- | ------- | ------- | ------- | ------- | 200 | 5 |

These results, about fixed effects, allow us to conclude about the importance of the individual analysis of the data for the different horticultural products against the global analysis and the importance of the factor location, because the cost of transport.

**Table 8. Results from the absolute convergence model for all vegetables export (percentage values relatively to the total of each country)**

|  | Const.[1] | Coef.[2] | F/Wald(mod.)[3] | F(Fe_OLS)[4] | Corr(u_i)[5] | F(Re_OLS)[6] | Hausman[7] | $R^2$[8] | N.O.[9] | N.I.[10] |
|---|---|---|---|---|---|---|---|---|---|---|
| FE[11] | 0.790* (15.180) | -1.114* (-24.020) | 576.820* | 3.390* | -0.915 | ------- | ------- | 0.563 | 641 | ------- |
| RE[12] | 0.196* (2.950) | -0.247* (-10.410) | 108.360* | ------- | ------- | 9.940* | 473.230* | 0.563 | 641 | ------- |
| OLS | ------- | ------- | ------- | ------- | ------- | ------- | ------- | ------- | ------- | ------- |
| DPD[13] | 1.521* (16.140) | -1.792* (-18.430) | 553.790* | ------- | ------- | ------- | ------- | ------- | 277 | 5 |

**Table 9. Results from the absolute convergence model for all vegetables import (percentage values relatively to the total of each year)**

|  | Const.[1] | Coef.[2] | F/Wald(mod.)[3] | F(Fe_OLS)[4] | Corr(u_i)[5] | F(Re_OLS)[6] | Hausman[7] | $R^2$ [8] | N.O.[9] | N.I.[10] |
|---|---|---|---|---|---|---|---|---|---|---|
| FE[11] | 0.134* (2.750) | -0.947* (-18.060) | 326.010* | 3.100* | -0.923 | ------- | ------- | 0.507 | 449 | ------- |
| RE[12] | -0.034 (-0.500) | -0.155* (-6.460) | 41.740* | ------- | ------- | 8.210* | 288.340* | 0.507 | 449 | ------- |
| OLS | ------- | ------- | ------- | ------- | ------- | ------- | ------- | ------- | ------- | ------- |
| DPD[13] | 0.520* (9.320) | -1.511* (-19.480) | 475.740* | ------- | ------- | ------- | ------- | ------- | 200 | 5 |

**Table 10. Results from the absolute convergence model for all vegetables export (percentage values relatively to the total of each year)**

|  | Const.[1] | Coef.[2] | F/Wald(mod.)[3] | F(Fe_OLS)[4] | Corr(u_i)[5] | F(Re_OLS)[6] | Hausman[7] | $R^2$ [8] | N.O.[9] | N.I.[10] |
|---|---|---|---|---|---|---|---|---|---|---|
| FE[11] | -0.347* (-7.450) | -1.074* (-23.080) | 532.690* | 3.550* | -0.935 | ------- | ------- | 0.544 | 641 | ------- |
| RE[12] | -0.038 (-0.450) | -0.253* (-10.350) | 107.140* | ------- | ------- | 15.880* | 429.780* | 0.544 | 641 | ------- |
| OLS | ------- | ------- | ------- | ------- | ------- | ------- | ------- | ------- | ------- | ------- |
| DPD[13] | -0.174* (-3.640) | -1.746* (-18.490) | 558.080* | ------- | ------- | ------- | ------- | ------- | 277 | 5 |

In the different results presented in the diverse tables (5 to 10) it is found evidence to the constant term which has always statistically significance and with values relatively high, what is a sign of lack of other variables in the model. In future works will be important try identify these variables.

## 5. Volatility analyze

It is analyzed, yet, the volatility of the data, taking into account tests like inverse chi-squared (P), Inverse normal and Inverse logit t.

All the results for all the variable, used in the estimations referred in the before section of this work, show that there is not stationary of the all data, what is preoccupant conclusion, because this show again that there is not a objective and coherent policy for the sector and for it international trade. As said the Keynesian theorist the external demand is the engine of the economy.

## 6. Conclusions

The horticultural sector is an important activity in Portugal, because the optimal natural conditions for this production. Has potential to growth and help the Portuguese in these days of national economic crises.

But it was not found a regularity of the import and export data for the vegetables sector, taking into account the results of the analysis of the data, and the observation of the stationary of the data with different tests.

In addition, Portugal import and export the same horticultural products from and for the same countries, namely the European countries.

As said by the new economic geography it was found some evidence of the importance of the costs of transport, considering the Portuguese partners in the transactions and the effects absorbed in fixed effects, namely from the cross-section data.

Portugal needs a coherent and adjusted policy to sector, with the objective of find some regularity the productions and in the international trade.